# Hydrogen Ion-Cut Single-Crystal Diamond Membranes for Wide and Ultrawide Bandgap Semiconductor Heterointegration

Xuanyu Zhou[1], Chenyu Wang[1], Dmitry Shinyavskiy[1], Lu Chen[1], Luke Suter[2], Andreas Graff[5], Michél Simon-Najasek[5], Vincent Gambin[3], Katherine T. Fountaine[3], Zhenqiang Ma[4], Frank Altmann[5], Matthias Muehle[2], Jung-Hun Seo[1,*]

[1]Department of Materials Design and Innovation, University at Buffalo, The State University of New York, Buffalo, NY 14260, USA

[2]Fraunhofer USA Inc., Center Midwest, East Lansing, MI 48824, USA

[3]Northrop Grumman Corporation, Redondo Beach, CA 90278, USA

[4]Department of Electrical and Computer Engineering, University of Wisconsin-Madison, Madison, Wisconsin, 53706, USA

[5]Fraunhofer Institute for Microstructure of Materials and Systems IMWS, Walter-Hülse-Str. 1, 06120 Halle, Germany

*Author to whom correspondence should be addressed. Electronic mail: junghuns@buffalo.edu

**Keywords:** diamond membrane, hydrogen implantation, ion-cut technique, WBG heterogeneous integration

## Abstract

Single-crystal diamond membranes (SCDms) were created using a hydrogen ion-cut process combined with a two-step post-implantation annealing, electrochemical etching, and micro-transfer printing. Monte Carlo simulations comparing H, He, C, and O implantation predicted a narrower vacancy-distribution profile for hydrogen under the implantation conditions examined. Raman spectroscopy was used to evaluate implantation-induced lattice damage and its recovery during annealing and membrane release, while optical bandgap analysis provided further evidence of restoration of the diamond lattice structure. After delamination and removal of residual graphitic material, freestanding SCDms with improved crystalline quality were obtained. The SCDms were subsequently transferred to GaN, SiC, AlGaN, and AlN substrates without an adhesive layer. The successful formation of diamond/WBG and diamond/UWBG heterostructures demonstrates the feasibility of integrating the SCDms with a range of semiconductor platforms. Overall, these results establish hydrogen implantation as an effective route for producing high-quality SCDms for future diamond-based heterogeneous integration applications.

## 1. Introduction

Single-crystal diamond (SCD) possesses an exceptional combination of material properties, including an ultrawide bandgap of approximately 5.5 eV, extremely high thermal conductivity exceeding 2,000 W $m^{-1}$ $K^{-1}$, high carrier mobility, and a large breakdown electric field exceeding 10 MV $cm^{-1}$, together with excellent chemical inertness and radiation hardness [1–3]. These properties make diamond one of the most promising semiconductor materials for next-generation high-power, high-frequency, and extreme-environment electronic and optoelectronic devices. In particular, its superior thermal conductivity enables efficient heat dissipation, which is critical for maintaining the performance and reliability of high-power devices. Diamond also exhibits excellent stability under harsh conditions, including high temperatures, intense radiation, and corrosive chemical environments, further broadening its applicability in advanced electronics and sensing technologies [1,2].

Despite these outstanding properties, the practical implementation of diamond in semiconductor devices remains limited by several key challenges. The availability of high-quality SCD substrates is restricted by their high production cost and relatively small wafer size compared with conventional semiconductor platforms such as Si, SiC, and GaN [4]. Furthermore, integrating diamond with other semiconductor materials is inherently difficult. Direct epitaxial growth of single-crystal diamond on conventional wide-bandgap (WBG) and ultrawide-bandgap (UWBG) semiconductors remains extremely challenging because of substantial lattice mismatch, differences in crystal symmetry, and incompatibility between the growth conditions required for diamond chemical vapor deposition (CVD) and those tolerated by the underlying semiconductors [5]. These fundamental material incompatibilities significantly hinder the realization of monolithically integrated diamond-based heterostructures. Nevertheless, the heterogeneous integration of diamond with WBG and UWBG semiconductor platforms, including GaN, AlGaN, SiC, AlN, and $Ga_2O_3$, has attracted considerable research interest in recent years [6-8]. Such hybrid structures can combine the desirable electronic properties of these semiconductors with the superior thermal-management capability of diamond, potentially improving device performance in high-power and high-frequency applications [2]. For example, diamond heat spreaders have been extensively investigated for GaN high-electron-mobility transistors (HEMTs) to mitigate self-heating and enhance power-handling capability. Previous studies have demonstrated the integration of polycrystalline diamond (PCD) films with GaN devices through CVD growth or wafer-bonding techniques to improve heat dissipation [9–10]. However, PCD contains grain boundaries and relatively high defect densities that can impede thermal transport and limit its suitability for active electronic integration.

To address these limitations, wafer bonding, surface-activated bonding, and ion-cut transfer techniques have been explored for the fabrication of diamond-based heterostructures [11-12]. Among these approaches, membrane-based semiconductor grafting has emerged as a promising strategy for the heterogeneous integration of materials with otherwise incompatible growth conditions [13]. In this approach, a thin crystalline semiconductor membrane is first created from a bulk or epitaxial substrate and subsequently grafted to a target substrate. This strategy has enabled the heterogeneous integration of single-crystalline semiconductor membranes of Si, Ge, III–V compounds, III-nitrides, SiC, and $Ga_2O_3$ with dissimilar materials while largely preserving their intrinsic properties [14-22].

However, the creation of single-crystal diamond membranes (SCDms) remains challenging because diamond cannot readily be grown heteroepitaxially on non-diamond substrates, making it difficult to incorporate conventional sacrificial layers for membrane release. In addition, the exceptional chemical and mechanical inertness of diamond further complicate membrane processing. In recent years, several studies have demonstrated the creation of SCDms using ion-cut techniques. Parikh et al. reported the lift-off of single-crystal diamond layers using high-energy $C^+$ or $O^+$ implantation to create a buried damaged layer, followed by thermal annealing and selective removal of the damaged region [23]. Also, Tran Thi et al investigated $O^+$ implantation-induced lift-off and demonstrated sub-nanometer surface roughness over tens of square micrometers, emphasizing the potential of an implantation-based approach for producing high-quality SCD surfaces [24]. Although these approaches successfully produced SCDms, heavy-ion implantation also introduces broad collision cascades and extensive lattice disorder with subsurface graphitization. The associated volume expansion and residual stress can cause the released membranes to bow or curl. Thus, successful lift-off alone does not establish that the released SCDm has the high crystalline quality, surface condition, and mechanical integrity required for heterogeneous integration. More recently, Guo et al. demonstrated direct bonding of SCDms onto various substrates using $He^+$ implantation, homoepitaxial overgrowth, and membrane transfer [25]. Their process removes the implantation-damaged diamond layer by plasma etching, retaining the overgrown layer as the final membrane. Although this strategy produces high-quality membranes, the additional growth and carrier-supported etching steps motivate a more direct route that prepares transferable membranes from the implanted diamond layer without post-implantation overgrowth.

As an alternative, $H^+$ implantation offers a promising route to SCDm synthesis with controlled implantation-induced damage. Its low ion mass may help limit collision-induced lattice disorder in the retained diamond, with implantation conditions selected to form a localized buried region for membrane separation. Combining this implantation strategy with post-implantation annealing and selective removal of damaged carbon enables the preparation of SCDms for heterogeneous integration without an additional diamond growth step. In fact, $H^+$-implantation-induced layer splitting has been widely employed to create single-crystalline semiconductor membranes from materials including Si, Ge, SiC, and $Ga_2O_3$. [17,26,27] Several studies have investigated these mechanisms in hydrogen-implanted diamond. Early work demonstrated hydrogen-induced layer splitting in diamond substrates, and subsequent studies reported ultrathin diamond-film lift-off and examined the formation of hydrogen-related platelets, bubbles, blisters, and exfoliated regions during annealing [28-31]. More recent SmartCut® studies combined hydrogen implantation with substrate bonding to transfer diamond films and evaluated the implantation damage retained within the transferred layers [32-33]. These investigations established the feasibility of hydrogen-assisted diamond separation but focused primarily on blistering mechanisms or the transfer of substrate-bonded diamond layers. The creation of lithographically patterned freestanding SCDm, removal of residual graphitic material from the released surface, and transfer-printing onto multiple functional WBG semiconductors remain comparatively underexplored.

In this work, we develop a process for the creation of high-quality SCDms using $H^+$-implantation and transferring them onto WBG and UWBG semiconductors by combining lithographic patterning to define the membrane geometry, $H^+$ implantation to create a localized buried damage region, and two-step post-implantation annealing to promote membrane delamination and recovery of crystalline quality. Subsequent electrochemical etching and backside plasma cleaning remove graphitized material and residual carbon from the membrane underside, preparing the released SCDms for transfering. Monte Carlo simulations compare the vacancy distributions produced by $H^+$, $He^+$, $C^+$, and $O^+$ implantation and evaluate the spatial confinement of their highly damaged regions. Cross-sectional scanning transmission electron microscopy (STEM) and electron energy-loss spectroscopy (EELS) examine the damaged carbon layer, while Raman and optical spectroscopy track changes in the diamond through implantation, two-step annealing, membrane release, and backside cleaning steps. These complementary analyses assess the initial damaged layer structure and the subsequent changes in crystalline and optical properties, connecting the processing sequence to the condition of the released SCDm. Finally, the released boron-doped SCDm are retrieved using a polydimethylsiloxane (PDMS) stamp and transferred to GaN, SiC, AlGaN, and AlN without using an adhesive layer to demonstrate proof-of-concept diamond/WBG and diamond/UWBG semiconductor heterostructures and establish the potential of this approach for heterogeneous integration.

## 2. Syntheiss of SCDm via hydrogen ion-cut process

**Figure 1(a)** schematically illustrates the generic synthesis process for the SCDm. A boron-doped SCD epitaxial layer stack, consisting of a 1.3 μm-thick p- layer ($4.5 \times 10^{17}$ cm$^{-3}$) [34] and a 300 nm-thick highly conductive p+ layer (~$3 \times 10^{20}$ cm$^{-3}$) [35] , was grown via microwave plasma chemical vapor deposition (MPCVD) on a commercial CVD diamond substrate from Excellent Diamond Products (EDP). (**Figure 1(a)(i)**). SCDm regions with various shapes were defined on the SCD substrate using a standard photolithography process, followed by deposition of a bilayer metal mask for subsequent plasma etching, as shown in **Figure S1(a)** of the Supplementary Information [36]. To obtain an SCDm with a target thickness of approximately 1.5 μm, masked regions on the diamond substrate were slightly over-etched by ~2 μm using an Inductively Coupled Plasma (ICP) etcher with $CF_4/O_2$ gases at 2/8 sccm (**Figure 1(a)(ii)**). After chemical removal of the metal mask using nichrome etchant (Sigma-Aldrich), the sample was mounted on a Cu plate (**Figure S1(b)**) and was implanted with high-dose, high-energy hydrogen ions at room temperature, using an implantation energy of 225 keV and a dose of ~$10^{17}$ cm$^{-2}$ at Michigan Ion Beam Lab.

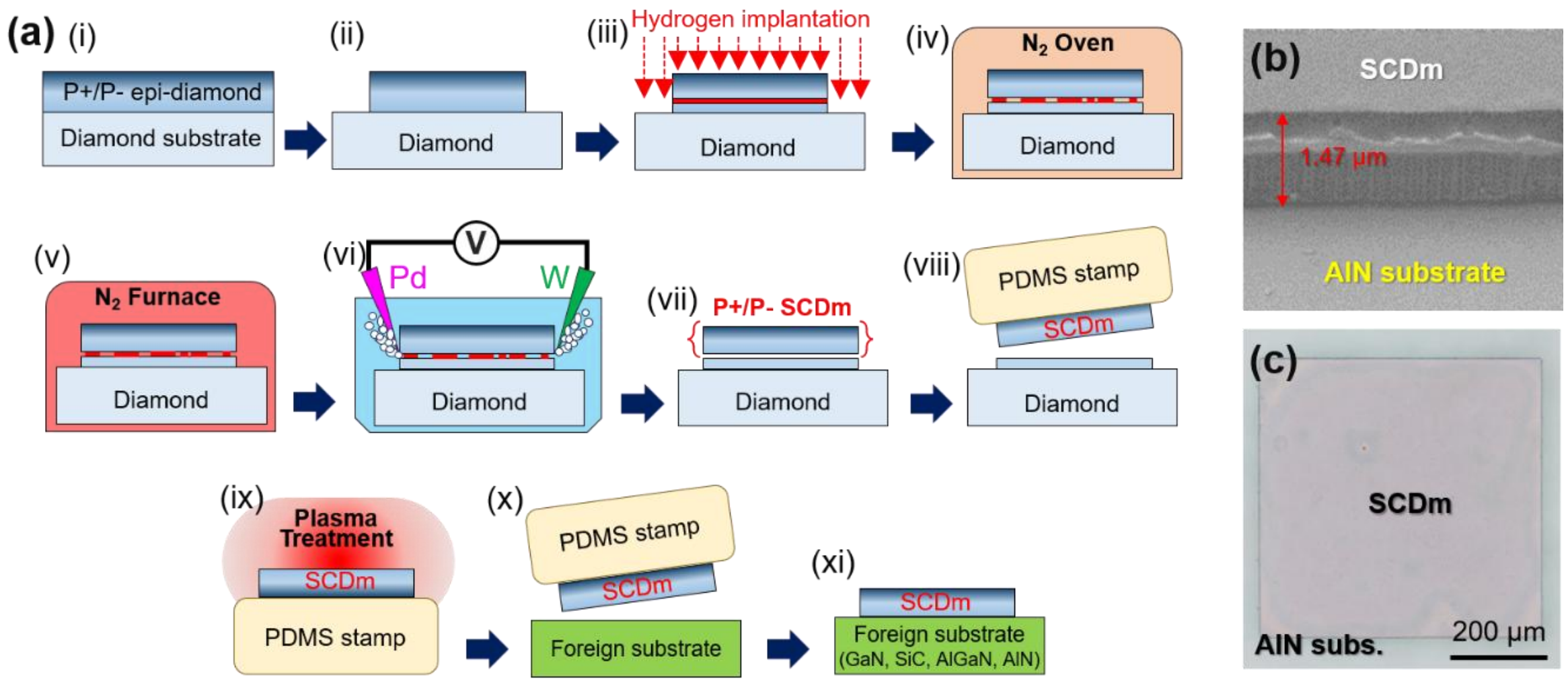

**Figure 1. (a)** Schematic illustration of the SCDm fabrication process. (i) diamond epitaxy growth, (ii) membrane patterning, (iii) H+ implantation, (iv) 1$^{st}$ low-temperature annealing, (v) high-temperature annealing, (vi)-(vii) electrochemical etching, (viii)-(ix) plasma etching on the bottom surface of SCDm (x)-(xi) micro-transfer printing. **(b)** Tilted-view SEM image and **(c)** microscopic image of an SCDm grafted to an AlN substrate.

(**Figure 1(a)(iii)**). The implantation conditions were selected based on the Monte Carlo simulation presented in **Figure 3** and **Figure S2(a)**. This simulation provides practical guidance for controlling the implantation depth and, consequently, the thickness of the SCDm over a broad range. Following hydrogen implantation, the samples underwent a two-step annealing process consisting of a low-temperature, long-duration anneal followed by a high-temperature, short-duration anneal (**Figure 1(a)(iv)** and **Figure 1(a)(v)**). The first annealing step was performed at 550 °C for 24 h to promote the formation and growth of hydrogen bubbles from the implanted hydrogen, while the second annealing step was conducted at 950 °C for 3.5 h to recover implantation-induced lattice damage. Both annealing steps were performed in an $N_2$ environment. **Figures S2(b)-(e)** show the evolution of the separation line from the as-implanted condition, 6, 18, and 24 hours at 550 °C, respectively. As shown in **Figure S3**, after the annealing step, approximately 80–90% of the membrane area was delaminated, which is denoted as "voids due to $H_2$ bubbles", while the remaining 10–20% remained attached to the substrate through an interfacial region that is identified in a later section as a "graphitic layer". To complete the membrane separation by removing the residual graphitic layer from the bottom surface of the SCDm, electrochemical etching was performed in deionized (DI) water at 30 V using tungsten and palladium needles, as shown in **Figure 1(a)(vi)** and **Figure S1(c)** of the Supplementary Information. After complete release of the membrane from the diamond substrate, the freestanding SCDm was retrieved using a polydimethylsiloxane (PDMS) stamp. With SCDm supported on the PDMS stamp, its exposed backside was plasma-etched to remove residual non-diamond carbon, as illustrated in **Figure 1(a)(ix)**. Then, the clean SCDm was transferred onto a foreign substrate by micro-transfer printing without using an adhesive layer (**Figure 1(a)(x)**). Finally, all heterostructures were annealed at 200 ºC for 5 min in $N_2$ ambient to enhance the bonding strength (**Figure 1(a)(xi)**). **Figure 1(b) and (c)** show representative tilted-view SEM and optical microscopic images of an SCDm transferred onto AlN, with an indicated membrane thickness of approximately 1.47 μm. In this study, the released SCDms were transferred onto GaN, SiC, AlGaN, and AlN substrates to demonstrate proof-of-concept diamond/WBG and diamond/UWBG semiconductor heterostructures, including diamond/GaN, diamond/SiC, diamond/AlGaN, and diamond/AlN. **Figure S4** of the Supplementary Information shows SCDm with different sizes and shapes and the capability to transfer them onto a foreign substrate.

The color variations observed in the microscopic images in **Figure S5** are attributed to two distinct mechanisms depending on their location within the SCD membrane. The color contours near the membrane edge in **Figure S5(a)** are primarily associated with thickness and topographical variations at the edge of the SCDm caused by localized over-etching during the electrochemical etching process. The reduced thickness near the SCDm edge is attributed to the formation of a thicker graphitic region caused by increased local hydrogen-ion exposure at the 7° implantation angle during $H^+$ implantation. Removal of this region during subsequent processing left the membrane edge several tens of nanometers thinner than the interior. This thickness nonuniformity alters the local optical interference condition and gives rise to the color contours. In contrast, the weak color variations observed within the interior region of the SCDm are likely related to residual graphite remaining on the bottom surface of the SCDm after the delamination process. As shown in **Figure S5(a)**, the graphitic residue locally modifies the optical response of the membrane/substrate structure, giving rise to the weak contrast variations observed within the interior region. However, as shown in **Figure S5(b)**, the interior of the SCDm exhibits a more uniform optical appearance after approximately 50 nm of plasma etching from the bottom surface of the SCDm (**Figure 1(a)(ix)**), suggesting that the interior color variations are associated with residual graphite.

### 3. Process design for hydrogen implantation

Various ion-cut-based methods for the creation of SCDms have previously been demonstrated using He, C, and O implantation [23-25]. **Figure 2(a)** compares the simulated vacancy-density profiles and the corresponding depth ranges of the implantation-modified layers for H, He, C, and O. In these simulations, the implantation conditions were selected to position the vacancy-density maxima at comparable depths of approximately 1.5 μm. The blue- and red-shaded regions denote the portions of each calculated vacancy profile exceeding 10% and 90% of its peak value, respectively. These relative thresholds are used to compare profile widths and do not represent experimentally established phase boundaries. As shown in **Figure 2(b)**, the overall damaged-layer thicknesses are 311, 338, 468, and 607 nm for H, He, C, and O implantation, respectively. More importantly, the corresponding high-density damage regions are only 32 nm for H but increase to 43, 81, and 118 nm for He, C, and O, respectively. This systematic increase in damage-zone width with ion mass can be attributed to the stronger nuclear stopping and broader collision cascades produced by heavier implanted ions. The calculated widths above the 90% threshold for C and O are approximately 2.5 and 3.7 times that for H, respectively. The thinner high-density damage region produced by H is also expected to minimize the loss of usable diamond thickness during membrane release and subsequent processing. It should be noted that a quantitative comparison of membrane stress and curvature among different implanted ions, however, is beyond the scope of the present study. Depending on the ion-cut mechanism, SCDm delamination may occur through hydrogen-induced platelet or bubble formation near the cleavage plane or through mechanical, chemical, or electrochemical removal of a graphitized sacrificial layer. The C and O profiles also exhibit substantial damage tails that extend farther into the surrounding diamond lattice and may be difficult to recover completely during annealing. Hydrogen implantation therefore produces a highly localized buried damage layer that is sufficiently concentrated to enable ion-cut release while minimizing disorder in the retained diamond. This spatial confinement facilitates the formation of a thin, well-defined graphitic sacrificial layer and suppresses the residual-stress gradients that can cause severe bending or bowing in SCDms created using C or O implantation. The improved membrane flatness enables more uniform and intimate contact with the foreign substrate during transfer, providing a critical advantage for adhesive-free heterogeneous integration.

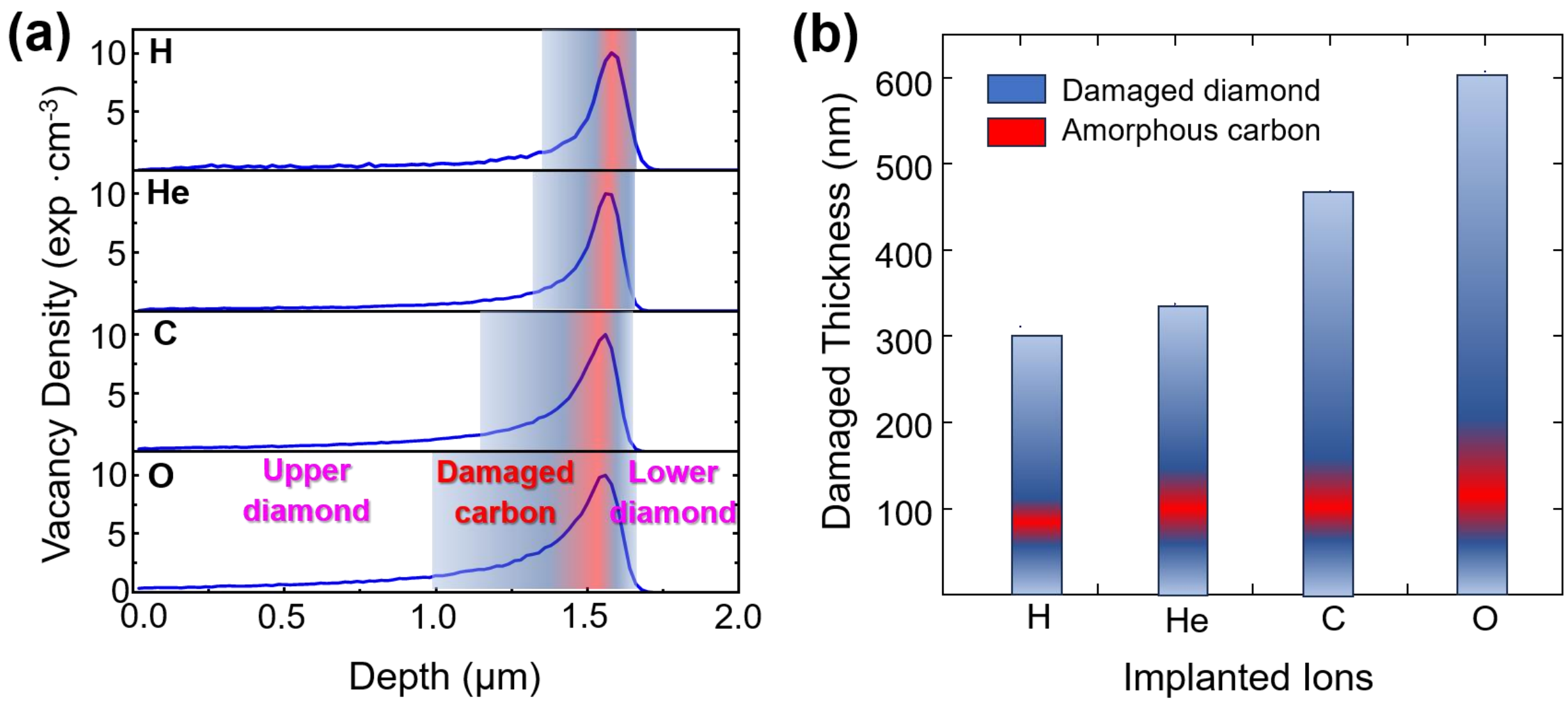

**Figure 2. (a)** Simulated depth profiles of implantation-induced vacancies in diamond with different implanted ions. The shaded regions indicate the highly damaged carbon layer, and the amorphous carbon layer. **(b)** Quantitative view of damaged carbon layer with different implanted ions.

Hydrogen implantation simulations were used to estimate the depth of the buried hydrogen-rich region and guide the selection of the target membrane thickness. **Figure 3(a)** shows the simulated relationship between hydrogen implantation energy and the predicted SCDm thickness at a fixed implantation dose of $1 \times 10^{17}$ cm$^{-2}$ using a Silvaco® TCAD simulator. The predicted thickness, estimated from the depth of the hydrogen-concentration maximum, increases monotonically with implantation energy. This relationship provides a means of selecting the approximate depth of the hydrogen-rich region and, consequently, the target membrane thickness. An implantation energy of 225 keV, denoted by the red star in **Figure 3(a)**, was selected to produce an approximately 1.5-µm-thick SCDm. **Figure 3(b)** and **(c)** present the corresponding two-dimensional hydrogen-concentration distribution and corresponding depth profile in diamond and predict a hydrogen-rich region approximately 1.5 µm below the surface. The depth profile reveals a hydrogen-concentration maximum at approximately 1.5 µm, with a peak concentration approaching $9 \times 10^{21}$ $cm^{-3}$. The hydrogen concentration decreases sharply on both sides of the peak, indicating a localized hydrogen distribution in the simulation.

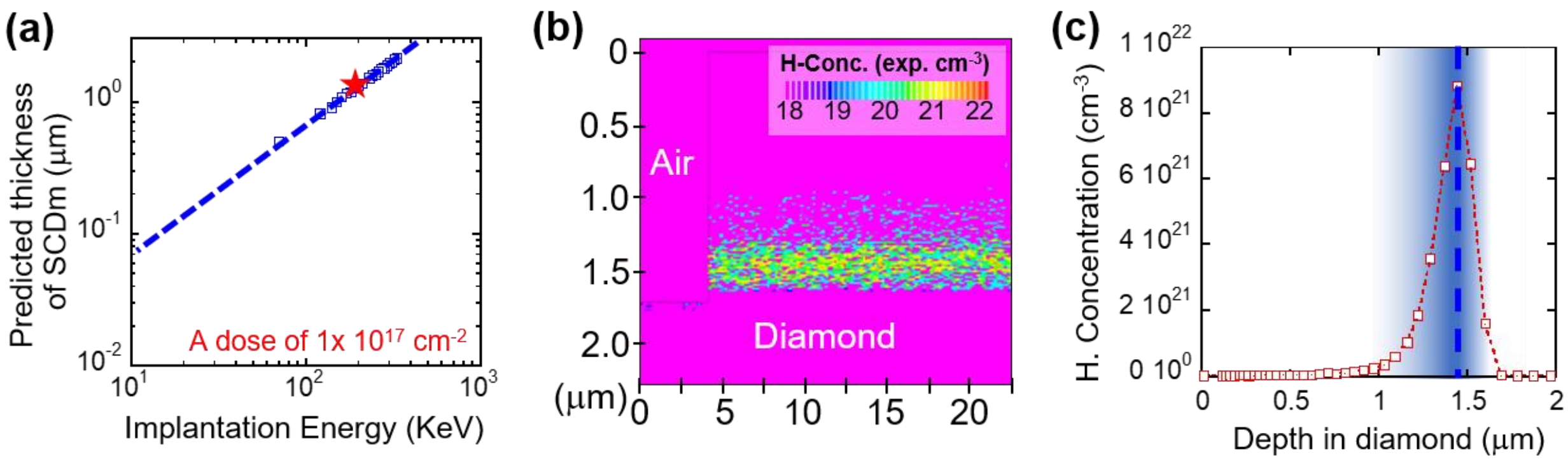

**Figure 3. (a)** Simulated SCDm thickness as a function of implantation energy at a fixed dose of $1 \times 10^{17}$ $cm^{-2}$, estimated from the depth of the hydrogen-concentration maximum. The red star denotes the selected implantation energy of 225 keV, corresponding to an SCDm thickness of approximately 1.5 µm. **(b)** TCAD two-dimensional simulation of hydrogen-concentration distribution in diamond at 225 keV. **(c)** Corresponding depth profile, showing a hydrogen-concentration maximum of approximately $9 \times 10^{21}$ $cm^{-3}$ at a depth of approximately 1.5 µm.

## 4. Structural and optical characterization of SCDm

Cross-sectional scanning transmission electron microscopy (STEM), fast Fourier transform (FFT) analysis, and electron energy-loss spectroscopy (EELS) were used to investigate the structural and bonding characteristics of the as-implanted diamond before thermal annealing. The low-magnification STEM image in **Figure 4(a)** reveals a laterally extended, buried damage band separating the upper diamond layer from the underlying substrate. Its subsurface location is qualitatively consistent with the predicted depth profile of implanted hydrogen shown in Figure 3. Higher-magnification images in **Figures 4(b) and 4(c)** resolve distinct regions of defective diamond and amorphous carbon within this band, indicating spatially disruption of the diamond lattice. The FFT patterns calculated from the corresponding regions further distinguish their local crystalline order. Discrete diffraction spots are observed in both the upper and lower diamond regions (**Figures 4(d)(i) and 4(d)(iv)**), demonstrating that lattice periodicity remains on both sides of the implantation-damaged band. In the defective diamond region, weak residual diffraction spots co-exist with substantial diffuse scattering (**Figure 4(d)(ii)**). This combination is consistent with a strongly disordered region that retains some crystalline order, rather than a fully amorphous structure. The amorphous carbon region, by comparison, exhibits a broad, diffuse halo without clearly resolved diffraction spots (**Figure 4(d)(iii)**), indicating the loss of long-range lattice periodicity. The diffuse scattering is consistent with short-range atomic correlations in a noncrystalline carbon network. These spatially resolved observations support a damaged region in which localized amorphization occurs adjacent to diamond that is disordered but retains crystalline order.

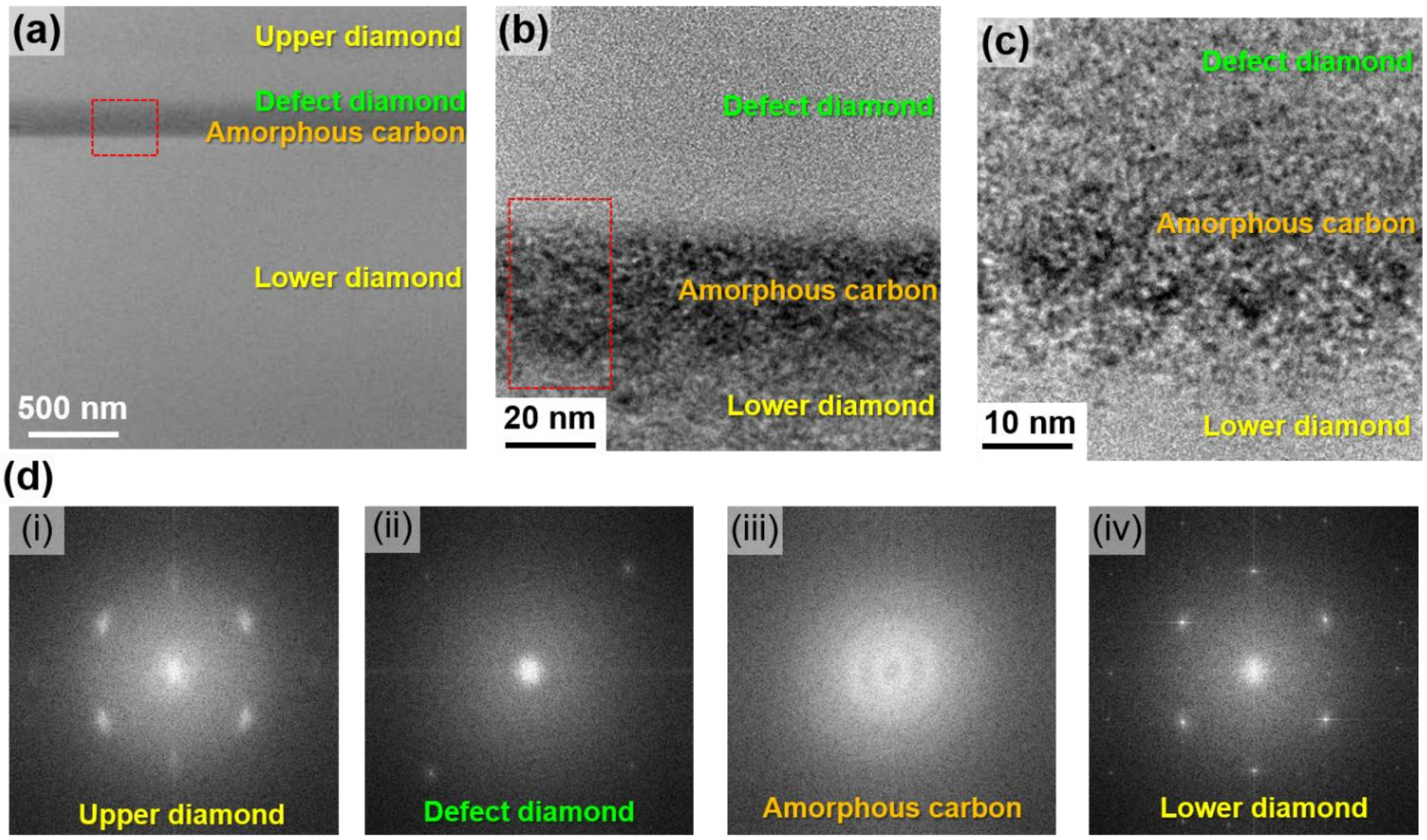

**Figure 4. (a)** Low-magnification STEM image showing the buried implantation-damaged layer between the upper and lower crystalline-diamond regions. **(b, c)** Higher-magnification STEM images revealing a defective crystalline-diamond region surrounding a highly disordered amorphous carbon region. The areas enclosed by the red dashed boxes in (a) and (b) are shown at higher magnification in (b) and (c), respectively. **(d)** FFT analysis of four regions of diamond.

Complementary information on the local bonding configuration is provided by the carbon K-edge EELS spectrum acquired from the amorphous region (**Figure S6)** in the Supporting Information). The near-edge feature at approximately 285–286 eV is attributed to the excitation of C 1s electrons into unoccupied $\pi^*$ states associated with $sp^2$-bonded carbon. Its presence supports the formation of $sp^2$ bonding within the implantation-damaged region and indicates a departure from the tetrahedral bonding characteristic of crystalline diamond. The higher-energy spectral intensity includes contributions from transitions into $\sigma^*$ states. Because $\sigma$ bonding occurs in both $sp^2$- and $sp^3$-coordinated carbon, this contribution alone cannot establish the presence or quantify the fraction of residual $sp^3$ bonding. The EELS and FFT results therefore provide complementary evidence: EELS identifies $sp^2$ bonding within the analyzed volume, while the diffuse FFT pattern establishes the absence of clearly resolved long-range crystalline order. Together, these signatures support the assignment of this region as $sp^2$-containing amorphous carbon.

The localized damage structure also provides a physical basis for interpreting the subsurface void formation and layer separation observed during the post-implantation annealing step. Implantation-induced defects and structurally disordered regions can provide trapping sites for implanted hydrogen. During annealing, thermally activated hydrogen redistribution and accumulation at these sites may promote the formation of hydrogen-containing cavities. In a hydrogen-pressure-assisted separation mechanism, molecular $H_2$ accumulates within these cavities, and the resulting internal pressure contributes to cavity expansion and crack opening along the damaged layer. Cavity growth and coalescence, together with the structural weakening of the implanted region, can then promote lateral crack propagation and partial delamination of the upper diamond layer. This interpretation is consistent with the annealing-dependent development of subsurface voids and separation, although the as-implanted STEM, FFT, and carbon K-edge EELS measurements cannot directly identify molecular $H_2$. The buried damaged band thus defines a preferential region for subsequent hydrogen-assisted separation, with removal of the remaining damaged carbon completing membrane release.

Micro-Raman spectroscopy was employed to monitor the structural evolution of the SCDm at each stage of the membrane fabrication process. The spectra were acquired using a Raman microscope (Renishaw inVia®) with a 514 nm excitation laser, an incident laser power of 25 mW, and a 50× objective lens. The analysis focused on the first-order diamond Raman line at approximately 1332 $cm^{-1}$ and the disorder-induced Raman bands associated with non-diamond $sp^2$-bonded carbon [37-39]. As shown in **Figure 5(a)(i)**, the as-grown diamond exhibits a well-defined first-order diamond peak centered at 1332 $cm^{-1}$ with a full width at half maximum (FWHM) of 9 $cm^{-1}$, confirming the high crystalline quality of the epitaxial diamond layer and providing a reference for the subsequent characterization stages. After hydrogen implantation, the Raman spectrum in **Figure 5(a)(ii)** exhibits substantial broadening of the diamond-related feature and the emergence of broad bands in the 1200–1350 $cm^{-1}$ and 1500–1600 $cm^{-1}$ regions. The broad lower-wavenumber feature is attributed to implantation-induced disorder within the $sp^3$-bonded carbon lattice, whereas the band at 1500–1600 $cm^{-1}$ is associated with the formation of a highly disordered, $sp^2$-rich carbon phase. [37] These spectral changes indicate extensive disruption of the diamond lattice and partial conversion of $sp^3$-bonded carbon into non-diamond carbon. After implantation, the measured optical transmittance decreased from approximately 80% to 65% in the visible wavelength range, consistent with increased optical losses associated with implantation damage. Raman spectrum acquired after the first low-temperature annealing step, shown in **Figure 5(a)(iii)** indicates partial recovery of the diamond lattice. The first-order diamond peak becomes more clearly resolved from the broad disorder-induced background, while the intensity of the $sp^2$-related band decreases. These changes indicate partial recovery of the $sp^3$-bonded lattice through the recombination of implantation-induced defects. After the second high-temperature annealing step, the Raman spectrum in **Figure 5(a)(iv)** shows further recovery of the crystalline diamond phase, as evidenced by the increase in the first-order diamond peak intensity and reduced linewidth. In addition, the $sp^2$ band shifts from approximately 1638 to 1597 $cm^{-1}$. This spectral evolution suggests a change in the bonding configuration and structural ordering of the implantation-induced $sp^2$-rich carbon phase and is consistent with its transformation toward a more graphitic structure. After release and backside cleaning, the spectrum is dominated by the diamond Raman peak with a FWHM of 13 $cm^{-1}$, compared with 9 $cm^{-1}$ for the as-grown sample. The suppressed $sp^2$-related signal is consistent with removal of non-diamond carbon, while the remaining linewidth difference indicates that the final Raman response is not identical to the as-grown reference (**Figure 5(a)(v)**). The substantial suppression of the $sp^2$-related Raman features demonstrates the effective removal of graphitic residues from the backside surface, while the pronounced diamond peak confirms the retention of the crystalline $sp^3$-bonded structure in the freestanding SCDm.

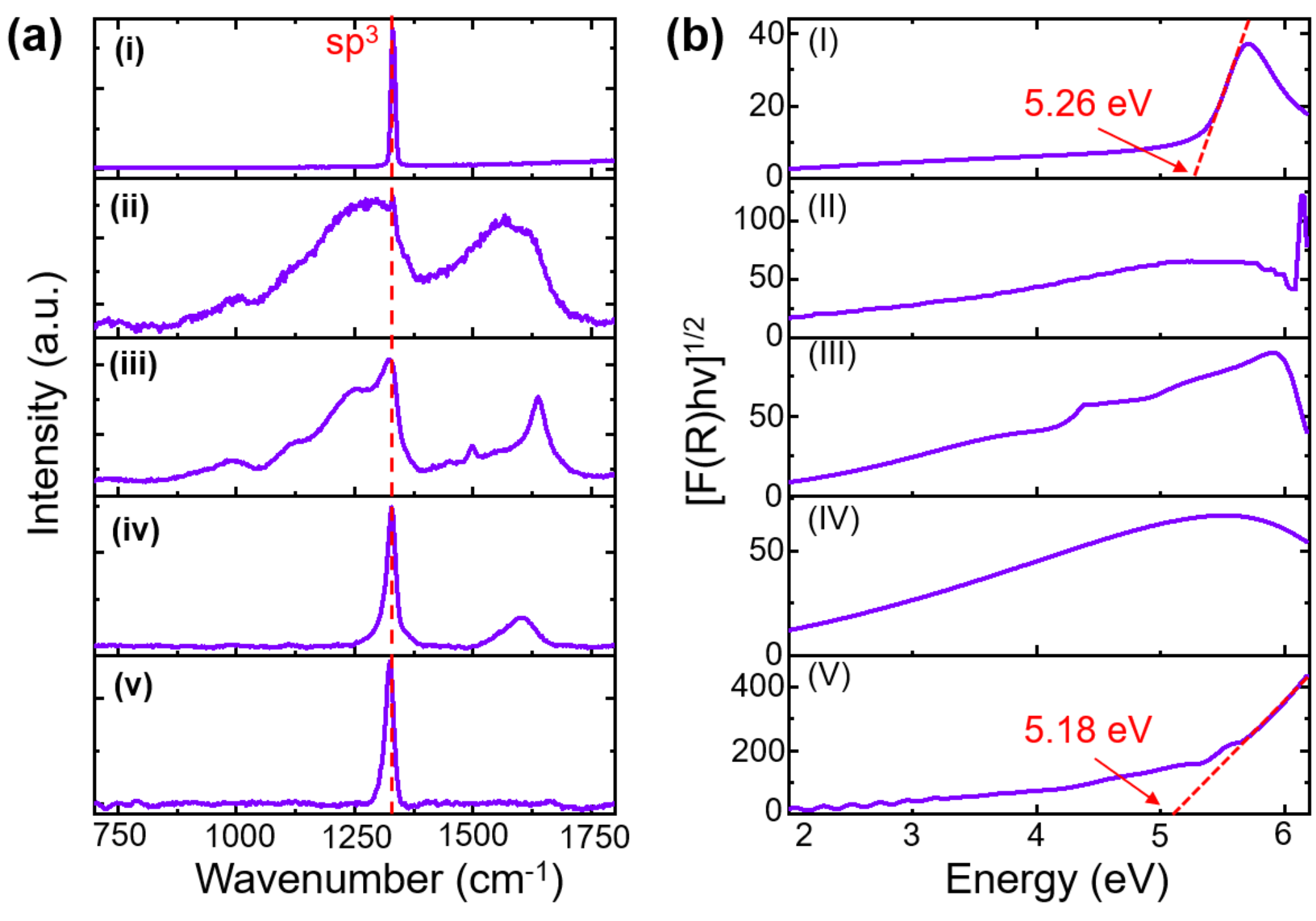

**Figure 5.** **(a)** Micro-Raman spectra and **(b)** corresponding Tauc plots obtained at successive processing stages: (i) as-grown diamond, (ii) after hydrogen implantation, (iii) after the first low-temperature annealing step, (iv) after the second high-temperature annealing step, and (v) after SCDm delamination and backside plasma cleaning. The red dashed vertical line in (a) marks the first-order diamond Raman peak at approximately 1332 $cm^{-1}$. The red dashed lines in (b) represent linear extrapolations used to estimate optical bandgaps of 5.26 and 5.18 eV for the as-grown diamond and the final SCDm, respectively.

In addition, characterization of the optical bandgap of the SCDm is important for the design of heterojunctions with other WBG semiconductors. **Figure 5(b)** shows the evolution of the optical bandgap throughout the SCDm fabrication process, determined from Tauc plots using the Kubelka–Munk transformation of the measured optical spectra [40]. As shown in **Figure 5(b)(i)**, the as-grown diamond exhibits an optical bandgap of approximately 5.26 eV, slightly lower than the intrinsic bandgap reported for high-purity diamond. [41] This difference is likely associated with boron-related impurity states and defect-induced band-tail absorption in the $p^+/p^-$ epitaxial layers. After hydrogen implantation and subsequent two-step post-implantation annealing processes, a reliable optical bandgap could not be extracted because the absorption edge became poorly defined, likely due to the combined contributions of implantation-induced lattice disorder and the buried graphitic layer formed during the ion-cut process. (**Figure 5(b)(ii)-(iv)**) After removal of the bottom graphitic layer by plasma etching, a well-defined absorption edge was recovered, yielding an optical bandgap of approximately 5.18 eV, which is close to the as-grown value of 5.26 eV as shown in **Figure 5(b)(v)**. The relatively small difference of 0.08 eV suggests substantial recovery of the implantation-induced lattice damage, while the remaining deviation may be associated with residual lattice disorder and defect-related electronic states in the SCDm.

## 5. Transfer of SCDm onto WBG Semiconductor Substrates

Finally, four representative WBG/UWBG heterostructures were fabricated through the SCDm transfer process: diamond/GaN, diamond/SiC, diamond/AlGaN, and diamond/AlN. These material combinations are of significant interest for future WBG/UWBG heterojunction devices, for which the electrical properties, interfacial quality, and carrier transport characteristics are important considerations. However, comprehensive characterization of the electrical properties, interface quality, and carrier transport behavior of each heterostructure would require a separate systematic investigation and is therefore beyond the scope of this study. Accordingly, the discussion here focuses primarily on demonstrating the successful formation and macroscale structural quality of the transferred WBG heterostructures, evaluated using Raman spectroscopy.

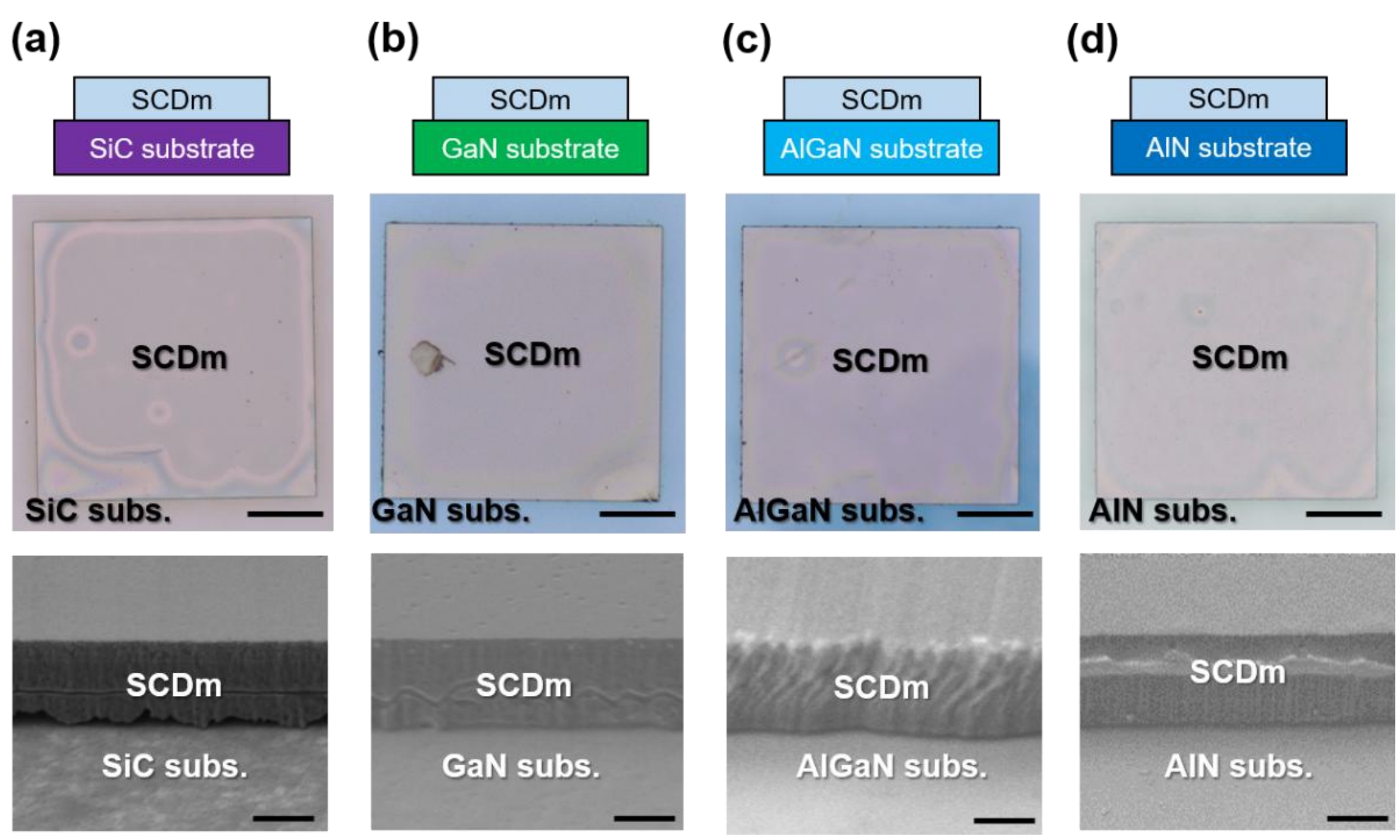

**Figure 6.** Heterogeneous integration of SCDm with WBG and UWBG semiconductor substrates. Schematic illustrations (top), microscopic images (middle), and SEM images (bottom) of SCDm transferred onto **(a)** SiC, **(b)** GaN, **(c)** AlGaN, and **(d)** AlN. The scale bar of optical microscopic images and SEM images are 200 μm, and 1 μm, respectively.

**Figure 6** presents schematic illustrations, optical microscopy images, and SEM images of SCDm transferred onto (a) SiC, (b) GaN, (c) AlGaN, and (d) AlN. The optical microscopy images show that the transferred SCDm, with lateral dimensions of 500 μm × 500 μm, remain intact on all four WBG and UWBG semiconductor substrates, with no obvious cracks or fractures. The shape and size of the membranes can be modified either before $H^+$ implantation or after transfer onto the target substrates. The SEM images provide complementary views of the SCDm-integrated heterostructures, showing that SCDm was integrated on top of the WBG and UWBG semiconductor substrates. Together, these observations demonstrate the successful physical integration of SCDm with the four semiconductor substrates while preserving the overall membrane geometry. **Figure 7** presents representative Raman spectra obtained from the four SCDm/WBG or UWBG semiconductor heterostructures. In all four structures, the characteristic diamond $sp^3$ phonon mode remains clearly identifiable near 1332 $cm^{-1}$, together with the characteristic Raman-active phonon modes of the corresponding WBG semiconductor. The presence of these distinct vibrational signatures indicates that the crystalline characteristics of both the transferred SCDm and the underlying semiconductor are maintained after heterostructure formation. In the SiC-based heterostructure, multiple characteristic acoustic and optical phonon modes, including $E_2$(TA), $E_2$(TO), $A_1$(LO), and folded transverse- and longitudinal-optical modes, are observed, consistent with the characteristic Raman response of hexagonal SiC. For the III-nitride-based heterostructures, the observed $E_2$(high) and $A_1$(LO) modes are consistent with the characteristic phonon modes of the corresponding wurtzite GaN, AlGaN, and AlN lattices. It should be noted that The spectra contain the characteristic Raman modes of diamond and the corresponding WBG substrates after transfer and annealing. These observations support the retention of the constituent crystalline phases and the absence of transfer-induced strain, but do not independently quantify the strain magnitude or interfacial bonding quality. Since the diamond $sp^3$ mode and several characteristic phonon modes of the WBG semiconductors are sensitive to lattice strain, the absence of measurable peak shifts suggests that the integration process does not introduce significant built-in strain into the heterostructures. In particular, the $E_2$(high) mode of wurtzite III-nitrides is highly sensitive to in-plane biaxial stress, while the position of the diamond $sp^3$ mode also depends on the stress state of the diamond lattice. The retention of the expected Raman peak positions therefore indicates that both the transferred SCDm and the underlying WBG or UWBG semiconductors preserve their lattice characteristics without appreciable transfer- or annealing-induced strain. Overall, these Raman results indicate that the SCDm transfer process maintains the crystalline integrity of both constituent materials without introducing substantial structural degradation. The results further support the feasibility of using transferred SCDm to form heterostructures with GaN, SiC, AlGaN, and AlN.

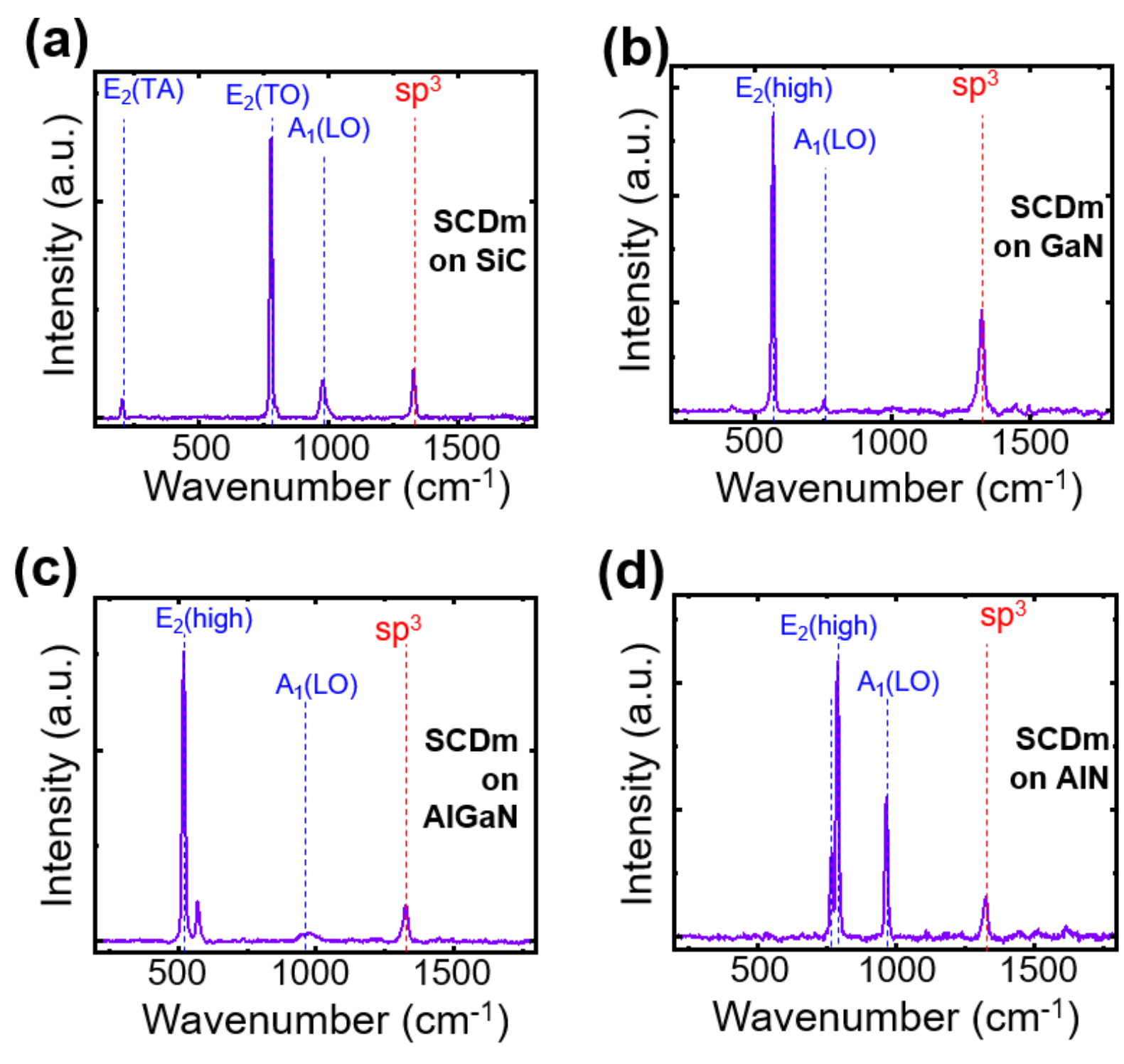

**Figure 7.** Representative Raman spectra of the SCDm/WBG and SCDm/UWBG semiconductor heterostructures: **(a)** SCDm/SiC, **(b)** SCDm/GaN, **(c)** SCDm/AlGaN, and **(d)** SCDm/AlN, respectively. The characteristic Raman-active phonon modes of each WBG or UWBG semiconductor are identified by blue dashed lines, while the red dashed lines indicate the first-order diamond $sp^3$ phonon mode near 1332 $cm^{-1}$.

## 6. Conclusion

In conclusion, we demonstrated a hydrogen ion-cut approach for the creation of SCDms and their heterogeneous integration with other WBG semiconductors. Monte Carlo simulations showed that hydrogen implantation produces a narrower and more localized vacancy distribution than He, C, and O implantation at a comparable projected depth, providing a favorable pathway for forming a buried damage layer while limiting implantation-induced disruption of the surrounding diamond lattice. The combination of hydrogen implantation, two-step post-implantation annealing, electrochemical release, and backside plasma cleaning enabled the creation of freestanding SCDms. Raman spectroscopy revealed substantial implantation-induced lattice disorder and $sp^2$-rich carbon formation after hydrogen implantation, followed by progressive recovery of the characteristic $sp^3$-bonded diamond structure during post-implantation processing. After membrane release and removal of the residual graphitic layer from the bottom surface of the SCDm, the SCDm exhibited a strong diamond Raman mode with well-suppressed $sp^2$-related features. Optical characterization further showed an optical bandgap of approximately 5.18 eV for the released SCDm, approaching the as-grown value of 5.26 eV and supporting substantial recovery of the optical characteristics of the diamond lattice. Finally, SCDms were successfully transferred to GaN, SiC, AlGaN, and AlN to form four representative diamond/WBG heterostructures without an adhesive layer. Raman spectra acquired after transfer and annealing contained the characteristic phonon signatures of both the diamond membranes and the underlying WBG semiconductors. Overall, these results establish hydrogen ion-cut SCDm fabrication and membrane transfer as a versatile route for integrating single-crystal diamond with dissimilar WBG semiconductor materials and provide a materials platform for future diamond-based heterogeneous electronic, optoelectronic, and thermal-management technologies.

## Acknowledgements

This work is supported by the Fraunhofer Program Affiliate Cooperation for Knowledge Transfer (PACT) program. This work is partially supported by the National Science Foundation under Award No. ECCS-2231026 and the Defense Advanced Research Projects Agency (DARPA) under contract 140D04-24-C-0061. The views, opinions and/or findings expressed are those of the authors and should not be interpreted as representing the official views or policies of DARPA or the U.S. Government.